\documentclass[11pt]{article}
\usepackage[dvipsnames,x11names]{xcolor}
\usepackage{epsfig}
\usepackage{lscape}
\usepackage{amssymb} 
\usepackage{graphicx}
\usepackage{footnote}
\usepackage[utf8]{inputenc}	
\usepackage{mathtools}
\usepackage{amsthm}
\usepackage{wasysym}
\usepackage{amsfonts}
\usepackage[english]{babel}
\usepackage{bigints}
\usepackage{mathrsfs}
\usepackage{lineno}
\usepackage{nicefrac}
\usepackage{enumerate}
\usepackage{appendix}
\usepackage{graphicx}
\usepackage{commath}
\usepackage{multirow}
\usepackage{tabu}

\usepackage{times}
\usepackage[T1]{fontenc}

\usepackage{algorithm} 
\usepackage{algpseudocode} 

\usepackage[most]{tcolorbox}
\usepackage{mdframed}

\usepackage{amsmath}
\usepackage{mathtools} 
\usepackage{mathrsfs}

\usepackage[]{natbib}
\usepackage{csquotes}

\usepackage{authblk}

\usepackage[colorlinks=true,citecolor=blue,linkcolor=black,urlcolor=blue]{hyperref}
\definecolor{light-gray}{gray}{0.91}

\usepackage[margin=2cm]{geometry}

\allowdisplaybreaks[1]
\usepackage{pifont}

\makeindex

\usepackage{sectsty}
\sectionfont{\color{BrickRed}}

\sectionfont{\color{DodgerBlue4}}
\subsectionfont{\color{DodgerBlue4}}
\subsubsectionfont{\color{DodgerBlue4}}
\paragraphfont{\color{DodgerBlue4}}

\newtheorem*{remark}{Remark}
\usepackage{authblk}

\newcommand{\imag}{\textbf{i}}

\newcommand{\error}{\mathcal{O}}

\def\Lm{\mathsf{L}}

\def\Linop{\mathcal{L}}
\def\adLinop{\mathcal{L}^{\dagger}}

\def\d{\textbf{d}}
\def\u{\textbf{u}}

\def\m{\textbf{m}}
\def\v{\textbf{v}}
\def\x{\textbf{x}}

\def\E{\textbf{E}}

\def\WF{\boldsymbol{\psi}}
\def\id{\mathbb{I}}
\def\error_op{\mathcal{E}}

\usepackage{color}

\begin{document}
	
\begin{flushleft}
	\flushleft
	\setlength{\baselineskip}{2.3em} 
	\textbf{\huge \color{DodgerBlue4}{Quantum observables as Fr\'echet sensitivity kernels}}\vspace{1ex}
	
	\vspace{1em}  
		
	\setlength{\baselineskip}{\normalbaselineskip} 
	
	\large{Rafael Abreu}\textsuperscript{1,*}\\[0.5em]  
	
	\scriptsize 
	\textsuperscript{1} \textit{Institut de Physique du Globe de Paris, CNRS, Université de Paris, Paris, France} \\[1em]  
	\textsuperscript{*} {email: rabreu@ipgp.fr}
	
	\normalsize 
\end{flushleft}

\begin{abstract}
We present a variational--adjoint interpretation of quantum mechanics in which the interaction between forward and adjoint wavefunctions defines a general sensitivity kernel of the system. This \emph{Fr\'echet interaction density} quantifies how perturbations in the wavefunction (or system parameters) influence a chosen observable. The familiar Born probability density appears as a special case when the adjoint wavefunction is chosen as the complex conjugate of the forward wavefunction, for which the interaction density becomes real and non-negative. Within this framework, probability is a particular positive-definite form of sensitivity described by the Fr\'echet interaction density.

In general, the variational--adjoint interpretation also produces Fr\'echet sensitivity kernels associated with other quantum observables, including momentum, energy, and spin. This suggests that the Born probability density belongs to a broader class of Fr\'echet sensitivity kernels associated with quantum observables. The proposed interpretation also provides a connection with time-symmetric interpretations of quantum mechanics and possible future applications in quantum control and quantum metrology.
\end{abstract}


\newpage

\section{Introduction}

The theory of quantum mechanics (QM) forms the theoretical foundation of many technological advances that define modern society. Far from being a mere abstract framework for understanding the microscopic world, QM continues to shape our world through cutting-edge technologies such as semiconductors, lasers, quantum computers, and magnetic resonance imaging \citep{nielsen2010quantum,dowling2003quantum,ballentine2014quantum}. These applications illustrate how the principles of QM have transitioned from the realm of theoretical physics to tangible innovations that underpin our everyday lives.

Yet, despite its success in predicting experimental outcomes with remarkable precision, the conceptual understanding of QM remains one of the great intellectual challenges of modern science. As Richard Feynman famously remarked, \textit{I think I can safely say that nobody understands quantum mechanics} \citep{Feynman1965}. The counter intuitive nature of the theory—manifested in phenomena such as superposition, entanglement, and wave–particle duality—continues to challenge our classical intuition about physical reality. This has motivated many attempts to better understand the meaning of the quantum theory, a task that is far from trivial.

Quantum mechanics is governed by the Schr\"odinger equation given by the following expression \citep[][]{griffiths2018introduction,shankar2012principles,zee2010quantum,sakurai2020modern}
\begin{align}
	\imag \, \hbar \, \partial_t \WF - \widehat{H}\WF = 0 ,
	\label{eq.Schroedinger_equation}
\end{align}
where $\imag$ is the imaginary unit, $\hbar$ is the reduced Planck's constant which determines how small the discrete units of energy, momentum, action and other quantities are at quantum scale. The symbol $\WF = \WF(\x, t)$ is the complex-valued wavefunction, which describes the quantum state of a system. The symbol $\widehat{H}$ is the Hamiltonian operator and it describes the total energy  (kinetic + potential) of the system. The symbol $\partial_t$ describes the first-order temporal derivative.

An interesting aspect of the Schr\"odinger equation is that the equation governs the deterministic evolution of $\WF$; however, the wavefunction $\WF$ itself does not provide a definite measurement outcome. It is the square modulus of the wavefunction, as prescribed by the Born rule, that gives the probability density for finding a quantum particle in a given state. In this sense, one of the fundamental aspects of quantum mechanics is the contrast between the deterministic evolution of the wavefunction and the probabilistic nature of measurement. The wavefunction provides the information from which the probabilities of the possible outcomes are obtained using the Born rule.

The Born rule relates the wavefunction $(\WF)$ to the probability of different measurement outcomes \citep{saunders2010many,neumaier2025born}. It is essential for understanding how we observe and interpret physical experiments using the Schr\"odinger equation. Specifically, it states that the probability \( P \) of measuring a particle at the location \( \x \) is given by the square modulus of the wave function itself,
\begin{align}
 P(\x,t_n) = |\WF(\x,t_n))|^2 = \WF(\x,t_n)\overline{\WF}(\x,t_n),
 \label{eq.Wave_Function_Collapse}
\end{align}
where the wave function $\WF$ is a complex valued vector and $\overline{\WF}$ is the complex conjugate. It is important to note that the wave function itself is complex, but the square modulus, \( |\WF|^2 \), gives a real and non-negative probability density for finding the particle at a specific location. The Born rule allows us to extract real-world information (like the position of a particle) from the wave function, which is inherently probabilistic.

The Born rule is critical in interpreting wave-particle duality. It tells us that while the wave function describes the wave-like evolution of a quantum system, the measurement process itself is discrete, and the outcomes are particle-like. The wave-like behavior is given \textit{before measurement}, where the quantum system is described by a superposition of all possible states. The wave function captures this wave-like behavior, encoding all the probabilities of different states the system could be in. The particle-like behavior is given \textit{after measurement}, when we measure a property of the system (e.g., position), the wave function "collapses" to a specific state, yielding a definite outcome. This collapse results in a particle-like behavior, where we observe the particle in a specific position or state. The Born rule thus provides the link between these two behaviors. 

The collapse of the wave function refers at what moment of time we compute the Born rule, but how do we decide that? The answer is simply related to when do we do the measurement or simply make an observation. After observation/measurement, the wave function chooses a particular state from the superposition (wave-like behavior), and the probability of observing this state is related to the square of the magnitude of the wave function at that point (Born rule, particle-like behavior). This collapse is non-deterministic and instantaneous, leading to a definite outcome. 

We can next pose the most natural question: what does it mean to make an observation? and as much incredible that it may sound, we still do not have an answer for that. This has lead to a large number of different interpretations of QM. Perhaps, the most famous is the Copenhagen interpretation \citep{bohr1935copenhagen}, which  provides a conceptual framework for understanding the measurement process in quantum mechanics. According to this interpretation the quantum system exists all the time in a superposition of all possible states (outcomes). It is after measurement that the wave function collapses to a definite outcome. The wave function is thus a mathematical tool for computing probabilities and it is not a description of a physical reality. The Copenhagen interpretation therefore highlights the fundamental role of the observer/measurement in determining the state of the system. 

Beyond Copenhagen, there exist many other interpretations of QM. While the Copenhagen interpretation emphasizes the probabilistic nature of measurement outcomes and the fundamental role of the observer \citep{bohr1935copenhagen}, others like the many-worlds interpretation \citep{Everett1957,DeWitt1970}, eliminates the need for wave-function collapse by postulating that all possible outcomes of a quantum event occur in branching, non-communicating universes. The Thermal Interpretation proposed by  \cite{neumaier2019coherent} assigns a direct physical meaning to quantum expectation values and their uncertainties. For \cite{neumaier2019coherent} an expectation value is interpreted as a property of the quantum system itself, and not as the statistical average obtained from an ensemble of repeated measurements.

Stochastic or hydrodynamic approaches offer alternative perspectives on the origin of quantum probabilities: Nelson's stochastic mechanics \citep{nelson1966derivation} derives quantum dynamics from an underlying classical stochastic process, while Madelung- or Bohmian-hydrodynamic formulations \citep{hall2002schrodinger,caticha2011entropic,yang2024quantum} recast the Schr\"odinger equation as coupled equations for a probability density and an action function, from which the wavefunction and the Born-rule probability naturally emerge. Recent measure-theoretic approaches, such as \cite{stoica2025born}, similarly attempt to explain the Born rule without postulating it, interpreting quantum probabilities as classical-like measures over a continuous underlying state space, and \cite{yang2024quantum} uses an information-based stochastic variational principle to derive the Schr\"odinger equation and Born probabilities from extremal properties of probability densities. Other reconstruction efforts motivate the Schr\"odinger equation from physical principles and assumed relationships between energy, momentum, and wave properties, incorporating the Born interpretation at the outset \citep{zhang2026schrodinger}.

We briefly review the major interpretations of QM in Table \ref{tb.qm_interpretations}. All these different interpretations aim to explain what the mathematical formalism of QM actually means — in other words, what reality is like if QM is true. While all interpretations agree on the predictive power of the theory, they differ in how they conceptually describe the nature of quantum states, measurement, and reality itself, highlighting the ongoing problem between the mathematical formalism of quantum mechanics and our philosophical understanding of physical reality.
\begin{table}
	\centering
	\small
	\renewcommand{\arraystretch}{1.4} 
	\setlength{\tabcolsep}{5pt}       
	\caption{Comparison of major interpretations of quantum mechanics. Each interpretation agrees on experimental predictions but differs in how it explains the role of the wavefunction, measurement, and reality.} \vspace{0.5cm}
	\begin{tabular}{|p{3cm}|p{5cm}|l|p{2.1cm}|p{2.2cm}|}
		\hline
		\textbf{QM interpretation} & \textbf{Core idea} & \textbf{Collapse?} & \textbf{Deterministic?} & \textbf{Refs.} \\
		\hline
		Copenhagen & Wavefunction encodes knowledge; collapse occurs upon measurement. & Yes & No & \cite{bohr1935copenhagen} \\
		\hline
		Many-Worlds & All possible outcomes occur in separate, branching universes; no collapse. & No & Yes &  \cite{Everett1957,DeWitt1970} \\
		\hline
		de Broglie--Bohm (Pilot-Wave) & Particles have definite positions guided by a pilot wave; deterministic but nonlocal. & No & Yes & \cite{Bohm1,Bohm2,Holland_1993} \\
		\hline
		Objective Collapse (GRW, Penrose) & Wavefunction collapse is a real physical process occurring spontaneously. & Yes & No & \cite{Ghirardi1986} \\
		\hline
		Consistent Histories & Quantum mechanics describes probabilities of entire histories, without collapse. & No & Yes &  \cite{griffiths1984consistent,griffiths2003consistent} \\
		\hline
		Quantum Bayes & Wavefunction represents personal belief; collapse is an update of information. & Epistemic & Subjective &  \cite{fuchs2014introduction} \\
		\hline
		Relational QM & Properties exist only relative to another system; no absolute state. & No & No & \cite{rovelli1996relational} \\
		\hline
		Transactional & Quantum events involve time-symmetric \textit{handshakes} between emitter and absorber. & No & Partly & \cite{cramer1986transactional,cramer2016quantum} \\
		\hline
		Two-Time (Two-State Vector) & System described by two wavefunctions — one evolving forward, one backward in time; outcomes depend on both past and future. & No & Yes  &  \cite{aharonov1964time} \\
			\hline
		Nelson's Stochastic Mechanics & Particle trajectories influenced by classical stochastic processes; Born rule arises naturally . & No & Yes  &  \cite{nelson1966derivation} \\
		\hline
		Madelung / Hydrodynamic Approaches & Schr\"odinger equation is rewritten in terms of probability density $\rho$ and action $S$. The wavefunction constructed as $\WF = \sqrt{\rho} e^{\imag S/\hbar}$, naturally yielding Born rule. & No & Yes &  \cite{hall2002schrodinger,caticha2011entropic,yang2024quantum} \\
		\hline
		Thermal (Neumaier) &
		Quantum quantities are described by their q-expectations and
		uncertainties, which represent objective properties of the system.&
		No &
		Yes &
		\cite{neumaier2019coherent} \\
		\hline
	\end{tabular}
	\label{tb.qm_interpretations}
\end{table}

\paragraph{The Contribution of this Work} 

We provide a different perspective by interpreting quantum probabilities using the \emph{Fr\'echet interaction density} derived from the variational structure of the Schr\"odinger equation. By introducing an adjoint wavefunction $\WF^\dagger$, whose source is determined by the chosen observable or measurement functional, the interaction between forward and adjoint wavefunctions describes the sensitivity of the system to perturbations in the wavefunction or system parameters.

The familiar Born rule appears as a \emph{particular case} within this interpretation. When the adjoint wavefunction is identified with the complex conjugate of the forward wavefunction, $\WF^\dagger=\overline{\WF}$, the resulting Fr\'echet interaction density is real, non-negative, and can be interpreted as a probability density after normalization. Other choices of the adjoint wavefunction give, in general, complex-valued sensitivity kernels associated with different observables.

In this sensitivity-based interpretation, probability arises from the interaction between the forward wavefunction and its particular adjoint choice. The formulation extends to quantum systems, potentials, and spatial dimensions for which the Hamiltonian admits the required adjoint structure under the adopted convolution bilinear form.

\section{Mathematical requisites}
\label{sec.notational_agreement}

We briefly review fundamental concepts of functional analysis used in this work. We give particular attention to the convolution bilinear form and the corresponding definition of the adjoint operator, since these concepts are fundamental to the variational formulation developed here. In particular, the time reversal contained in the convolution bilinear form is essential for understanding the definition of the adjoint wavefunction and the Fr\'echet interaction density introduced later. It is crucial to note that this distinction is not merely notation: using the usual inner-product adjoint instead of the convolutional adjoint leads to a different treatment of the first-order derivatives and therefore to a different adjoint formulation and interpretation. For these reasons, we introduce these concepts explicitly before developing the mathematical formulation. For a more detailed introduction we refer to \cite{zeidler1984nonlinear,zeidler1990nonlinear,griffel2002applied}.

\subsection{Inner products}

We denote by the inner product of two scalar $u(x),v(x)$ and two vector $\u(\x),\v(\x)$ functions, as follows
\begin{align}
	\left\langle u,v \right\rangle_x = \int_x u \, v \dif x, \qquad \left\langle \u,\v \right\rangle_x =  \int_{\Omega} \v^T \u \dif \x =  \int_{\Omega} \u \cdot \v \dif \x = \int_{\Omega} u_i \, v_i \dif \x,
\end{align}
respectively. Analogously, the inner product over time and space of two scalars $u(x,t),v(x,t)$ and two vectors $\u(\x,t),\v(\x,t)$ functions, as follows
\begin{align}
	\left\langle u,v \right\rangle_{x,t} = \int_x \int_T u \, v \dif t \dif x , \qquad \left\langle \u,\v \right\rangle_{x,t} = \int_{\Omega} \int_T \v^T \u \dif t \dif \x =  \int_{\Omega} \int_T \u \cdot \v \dif t \dif \x = \int_{\Omega} \int_T  u_i \, v_i \dif t \dif \x,
\end{align}
respectively. 

A symmetric matrix is defined as a matrix $A$ such that $A^T=A$, where $A^T$ refers to the transposed matrix defined by $A^T_{ij}=A_{ji}$. Then it follows that for any $x,y\in \mathbb{R}^n$,
\begin{align}
	\left\langle x,Ay \right\rangle = \left\langle A^Tx,y \right\rangle .
	\label{eq.Matrix_adjoint}
\end{align} 

\subsection{Linear operators, their adjoints and convolutional adjoints}

Let $N,M$ be vector spaces. An operator $\Linop:N\to M$ is linear if 
\begin{align}
	\Linop(ax+by) = a \Linop x + b \Linop y ,
\end{align}
for all scalar $a,b$ and all $x,y\in N$. The adjoint  of a linear operator comes from a generalization of the matrix inner product given in eq. \eqref{eq.Matrix_adjoint}. Let $\Linop:\mathcal{H}\to \mathcal{H}$ be a bounded linear operator on a Hilbert space $\mathcal{H}$, then there is a unique operator $\Linop^{\dagger}:\mathcal{H}\to \mathcal{H}$ such that
\begin{align}
	\left\langle x,\Linop^{\dagger}y \right\rangle = \left\langle \Linop x,y \right\rangle \qquad \text{for all} \quad x,y\in \mathcal{H} .
\end{align}
The linear and bounded operator $\adLinop$ is called the adjoint operator of $\Linop$. 

A linear bounded operator $\Linop:N\to M$ is invertible if for each $x\in M$ there is one and only one $y\in N$ such that $\Linop y = x$. The mapping $x \mapsto y$ is called the inverse of $\Linop$ and we denote it by $y=\Linop^{-1}x$. The adjoint operator $\adLinop$ satisfies \citep{griffel2002applied}
\begin{align}
	\left(\Linop^{\dagger} \right)^{\dagger}=\Linop, \qquad  \left(\Linop^{\dagger} \right)^{-1} = \left(\Linop^{-1} \right)^{\dagger} , \qquad \left(\Linop_1 \Linop_2 \right)^{\dagger} = \left(\Linop_2\right)^{\dagger} \left(\Linop_1 \right)^{\dagger} ,
\end{align}
where $\Linop_1,\Linop_2$ refer to two different linear bounded operators. The adjoint operator is called self-adjoint (or Hermitian) if $\adLinop=\Linop$.

There are two key points to remember about the adjoint operators previously defined:
\begin{enumerate}
	\item The adjoint operator \( A^{\dagger} \) is typically defined with respect to an inner product space.
	\item The adjoint satisfies the condition that the inner products \( \langle A x, y \rangle = \langle x, A^{\dagger} y \rangle, \forall x , y \).
\end{enumerate}

In the context of a Hilbert space \( \mathcal{H} \), the convolution of two functions \( f \) and \( g \) is defined as the inner product between \( f \) and the shifted version of \( g \) as follows:
\begin{align}
	(f \star g)(x) = \int_{-\infty}^{\infty} f(\tau)g(x-\tau) \dif \tau = \langle f(\cdot), g(x - \cdot) \rangle = \langle f , g \rangle^c ,
	\label{eq.convolution}
\end{align}
where \( x \) is the point at which the convolution is evaluated, and \( g(x - \cdot) \) represents the shifted function \( g \), with \( x \) shifting the argument of \( g \). 

For the finite time interval considered in this work, the convolution bilinear form evaluated at time $T$ can be written as
\begin{align}
	\left\langle f,g\right\rangle_T^c
	=
	\int_0^T f(t)\,g(T-t)\,\dif t
	=
	\int_0^T f(T-t)\,g(t)\,\dif t .
	\label{eq.convolution_bilinear_time}
\end{align}
It is crucial to note, for this work, that the convolution bilinear form involves the time reversal of one of the two functions. Which function is regarded as time reversed is only a matter of representation. 

The Adjoint operator in terms of the convolution bilinear form is defined as follows: Let \( A \) be a bounded linear operator acting on a Hilbert space \( \mathcal{H} \). The adjoint operator \( A^{\dagger} \) of \( A \) with respect to the convolution bilinear form satisfies the following equation for all \( x, y \in \mathcal{H} \):
\begin{align}
	\langle A x, y \rangle = \langle x, A^{\dagger} y \rangle ^c, \quad \forall x , y \in \mathcal{H},
\end{align}
where \( \langle \cdot, \cdot \rangle^c \) denotes the convolution on the Hilbert space \( \mathcal{H} \). The operator \( A^{\dagger} \) is the unique operator that satisfies this relation, provided that \( A \) is a bounded operator \citep{tonti1973variational}.  

The important difference with the usual adjoint definition is that one of the functions in the bilinear form is evaluated at the time-reversed argument $T-t$, as explicitly shown in eq. \eqref{eq.convolution_bilinear_time}.

\begin{remark} It is important to note that the convolution bilinear form used here does not require the functions involved to be real valued. For complex-valued functions, the resulting bilinear form, and consequently the corresponding Fr\'echet sensitivity kernels, can in general be complex valued.
\end{remark}

\section{A variational–adjoint formulation of the Schr\"odinger equation}

For any experiment in QM involving the Schr\"odinger equation, we define the general error or misfit functional $\E$ as follows 
\begin{align}
	\E & = \error_op (\textbf{d} ,\WF),
	\label{eq.quantum_chi_squared}
\end{align} 
where $\error_op$ represents an operator that computes the discrepancy between observations $\textbf{d}$ and predictions (using the wave function $\WF$). The Schr\"odinger equation, in operator form, can be written as follows
\begin{align}
	\Linop (\m) \WF = \imag \, \hbar \, \partial_t \WF - \widehat{H}\WF = 0 ,
	\label{eq.Schroedinger_equation_Operator}
\end{align}
where $\hbar$ is the reduced Planck's constant, $\m$ are the model parameters, $\WF = \WF(\x, t)$ is the wavefunction and $\widehat{H}$ is the Hamiltonian operator. Since $\WF\in\mathbb{C}$, the variational--adjoint formulation developed below is, in general, complex valued. We therefore use the convolution bilinear form introduced in Sec.~\ref{sec.notational_agreement}, without introducing complex conjugation unless it is explicitly specified.

\subsection{Lagrange optimization}

We apply the method of Lagrange multipliers to minimize the error function $\E$ (eq. \eqref{eq.quantum_chi_squared}) subject to the constraint that the observations can be properly described by the Schr\"odinger operator $\Linop(\m) \WF$ (eq. \eqref{eq.Schroedinger_equation_Operator}). This constrained optimization problem can be formulated as follows
\begin{align}
	\chi = \int_{0}^{T} \int_{\Omega}  \left[\error_op (\textbf{d} ,\WF) - \Lm \left(\Linop(\m) \WF \right)\right] \dif \x \dif t ,
	\label{eq.Adjoint_General_Functional}
\end{align}
where \( \Omega \) is the spatial domain, and \( \Lm \) is a vector-valued Lagrange multiplier that enforces the physical constraint. Using the convolution bilinear form (see \citep{tonti1973variational,abreu2024understanding}), we can express eq. \eqref{eq.Adjoint_General_Functional} in a more compact operator form:
\begin{align}
	\chi = \left\langle \id, \error_op (\textbf{d} ,\WF) \right\rangle_{\x,t}  - \left\langle \Lm ,\Linop(\m) \WF \right\rangle_{\x,t} ,
	\label{eq.QM_operator}
\end{align}
where $\id$ is the identity operator and  $\left\langle \right\rangle$ refers ot the inner product. The Karush–Kuhn–Tucker (KKT) conditions are a set of necessary conditions for a solution in a constrained optimization problem to be optimal \citep{hanson1981sufficiency,boyd2004convex,hanson1999invexity} which must be satisfied, leading to the following explicit expressions 
\begin{align}
	\begin{aligned}
		\partial_\m \chi = \left\langle \Lm,\delta_{\m} \Linop \WF \right\rangle^c_{\x,t} = 0, \qquad \partial_\Lm \chi = \Linop \WF  = 0, \qquad \partial_{\WF} \chi = \partial_{\WF} \error_op + \Linop^{\dagger} \Lm = 0 , \qquad \partial_\d \chi = 0 ,
		\label{eq.KKT_conditions}
	\end{aligned}
\end{align}
where the upper dagger $(\dagger)$ refers to the adjoint operator with respect to the convolution bilinear form (see \cite{tonti1973variational,abreu2024understanding}). Note that the third term in eqs. \eqref{eq.KKT_conditions}, naturally leads to the adjoint equation
\begin{align}
	\underbrace{\Linop^{\dagger} \WF^{\dagger}}_{\text{adjoint equation}} = \underbrace{\partial_{\WF} \error_op}_{\text{source term}} ,
	\label{eq.General_Adjoint_Equation}
\end{align}
where we identify the time-reversed Lagrange multiplier with the adjoint wavefunction $\WF^{\dagger}$ in order to satisfy the boundary conditions (see \cite{abreu2024understanding,abreu2025frechetrootkernelcertain}).

It is crucial to note that the time reversal associated with the convolution bilinear form is therefore already included in $\WF^{\dagger}$. More explicitly, the
convolution pairs the forward wavefunction at time $t$ with the adjoint wavefunction at time $T-t$. We can define
\begin{align}
	\Lm(\x,T-t) = \WF^{\dagger}(\x,t) .
	\label{eq.Lagrange_Adjoint_Defition}
\end{align}
Thus, $\WF^{\dagger}(\x,t)$ denotes the time-reversed adjoint wavefunction, so that both forward and adjoint wavefunctions are evaluated at the same physical time $t$.

Note also that the time-reversed adjoint wavefunction $\WF^{\dagger}$ is governed by the same Schr\"odinger equation ($\Linop^{\dagger}$ in eq. \eqref{eq.General_Adjoint_Equation}),
since the Hamiltonian operator is self-adjoint \citep{sakurai2020modern,griffiths2018introduction} and the term $\imag\,\hbar\,\partial_t\WF$ is also self-adjoint with respect to the convolution bilinear form (see \cite{abreu2024understanding} for further details).

Note that imposing $\partial_\Lm \chi = \partial_{\WF} \chi = 0$ implies that $ \left\langle \Lm ,\Linop(\m) \WF  \right\rangle^c_{\x,t}=0$, which allow us to write 
\begin{align}
	\partial_\m \chi=\partial_\m \E =\left\langle \WF^{\dagger},\delta_{\m} \Linop \WF \right\rangle^c_{\x,t} ,
	\label{eq.Quantum_Frechet_derivative}
\end{align}
which can be understood as the Fr\'echet derivative of the error or misfit function $\E$ (eq. \eqref{eq.quantum_chi_squared}). Note that the term $\delta_{\m} \Linop \WF$ is simply the derivative with respect to the model parameters of the Schr\"odinger equation. 

We can write eq. \eqref{eq.Quantum_Frechet_derivative} as follows
\begin{align}
	\partial_\m \chi=\partial_\m \E =\left\langle \delta_{\m} \Linop^{\dagger}  \WF^{\dagger},\WF \right\rangle^c_{\x,t} .
	\label{eq.Quantum_Frechet_derivative2}
\end{align}

\subsection{ The Fr\'echet interaction density}

We define the Fr\'echet interaction density as follows
\begin{align}
	K_R =  \WF \, \WF^\dagger  ,
	\label{eq.Frechet_interaction}
\end{align}
which represents the interaction between the forward and adjoint wavefunctions in eq. \eqref{eq.Quantum_Frechet_derivative2}, and its corresponding time integrated kernel
\begin{align}
	\mathsf{K}_R    =\int_T K_R \, \dif t .
\end{align} 

Recall that the time reversal associated with the convolution bilinear form is already included in $\WF^{\dagger}$, as defined in eq. \eqref{eq.Lagrange_Adjoint_Defition}. Thus,
$\WF^{\dagger}(\x,t)$ and $\WF(\x,t)$ are evaluated at the same physical time $t$. Note that, in general, eq. \eqref{eq.Frechet_interaction} defines a sensitivity or influence field that is not necessarily positive definite.

Fr\'echet sensitivity kernels given in eq. \eqref{eq.Quantum_Frechet_derivative2} can then be written in the general form as follows
\begin{align}
	K_\m  =  \left\langle \Linop^{\dagger}_\m \left[\WF^\dagger\right] ,\WF  \right \rangle^c_{t} .
	\label{eq.General_Frechet_Kernel}
\end{align}

Thus the interaction density $K_R(x,t)$ (eq. \eqref{eq.Frechet_interaction}) works as a fundamental bilinear quantity, the instantaneous overlap between forward and adjoint wavefields, from which the Fr\'echet kernels are constructed, and the operators $\mathcal{L}_\m$ encode how the governing physics weights this interaction before the time integration. 

Note that the physically relevant components of each kernel $\Linop^{\dagger}_\m$ corresponds to the derivatives of the wavefield that enter the governing equation. This guaranties that the sensitivity reflects how model parameters influence the physics of the PDE in study.

Note that in general, the Fr\'echet interaction density $\WF^{\dagger}\WF$ may be complex-valued. If observations $\d$ are well enough described by the Schr\"odinger equations and that the adjoint wavefield $\WF^{\dagger}$ satisfies eq. \eqref{eq.General_Adjoint_Equation}, the Fr\'echet interaction density encodes how different regions of space--time contribute to the variation of the functional $\chi$.
	
A direct probabilistic interpretation arises for a particular and physically motivated choice of the adjoint field. If one identifies the adjoint wavefield with the complex conjugate of the forward wavefunction, $\WF^{\dagger}=\overline{\WF}$, which can be obtained by selecting an appropriate adjoint source term in eq.~\eqref{eq.General_Adjoint_Equation}, then the Fr\'echet interaction density reduces to
\begin{align}
	\WF^{\dagger}\WF = \overline{\WF}\WF = |\WF|^2 ,
\end{align}
which is real and non-negative by construction. In this case, the Fr\'echet interaction density can be interpreted, after normalization, as describing the probability density of the event. In this case, the Fr\'echet interaction density coincides with the standard Born probability density. 

This suggests how the standard quantum probability structure emerges as a specific realization of the more general sensitivity kernel defined by the variational formulation. While the Fr\'echet interaction density $\WF^{\dagger}\WF$ represents a sensitivity or influence field, the identification $\WF^{\dagger}=\overline{\WF}$ provides the minimal structure required to recover a positive-definite quantity consistent with the probabilistic interpretation of quantum mechanics. 

\subsection{Example: the 1D Schr\"odinger equation} 

The next 1D example constitutes an explicit illustration of the general framework: the quantity \(\psi^{\dagger}\psi\) defines a sensitivity field, and the familiar probabilistic structure of quantum mechanics emerges when the adjoint field is identified with the complex conjugate of the forward wavefunction.

In one spatial dimension, the time-dependent Schr\"odinger equation is governed by the Hamiltonian operator
\begin{align}
	\hat{H} \;=\; -\frac{\hbar^2}{2m}\,\partial_x^2 + V(x),
\end{align}
where \(m\) is the particle mass and \(V(x)\) the potential energy. The corresponding Schr\"odinger equation reads
\begin{align}
	\imag \hbar\, \partial_t \WF(x,t)
	\;=\;
	\left[
	-\frac{\hbar^2}{2m}\,\partial_x^2 + V(x)
	\right]\WF(x,t).
\end{align}
Different choices of \(V(x)\) generate different dynamical regimes, but the variational structure derived above applies in all cases. The constrained Lagrange minimization problem eq. \eqref{eq.Adjoint_General_Functional} can thus be written as follows
\begin{align}
	\chi =   \frac{1}{2} \left \langle  (\d  - \WF), (\d  - \WF) \right \rangle_{x,t} -  \left \langle \Lm , \left[ \imag \, \hbar \, \partial_t \WF +\frac{\hbar^2}{2m} \partial^2_x \WF -  V \WF \right] \right \rangle_{x,t}^c ,
	\label{eq.1D_Schroedinger_Constrained_Action}
\end{align}
where the vector Lagrange multiplier $\Lm(x,t)$ (which minimizes the functional $\chi$) remains to be determined. 

Taking the total variation of the $\chi$ functional given in eq. \eqref{eq.1D_Schroedinger_Constrained_Action}, we can write the following
\begin{align}
	\begin{aligned}
		\delta_\m \chi = &  \left \langle  \Lm , \left[\imag \, \delta \hbar \, \partial_t \WF + \frac{\hbar^2}{2m^2} \delta m \, \partial^2_x \WF \right] \right \rangle_{x,t}^c  ,\\
		\delta_{\psi} \chi = & \left \langle (\d  - \WF) , \delta \WF  \right \rangle_{x,t}  - \left \langle  \Lm , \left[ \imag \, \hbar \, \partial_t \delta \WF +\frac{\hbar^2}{2m} \partial^2_x \delta \WF -  V \delta \WF \right] \right \rangle_{x,t}^c .
	\end{aligned}
	\label{eq.1D_General_Complex_Schroedinger_Constrained_Action}
\end{align}

In the absence of perturbations in the model parameters $(\delta m = 0)$, the stationary condition $(\delta_\m \chi + \delta_{\Lm} \chi + \delta_{\psi} \chi =0)$, leads to 
\begin{flalign}
	\imag \, \hbar \,  \partial_t \WF^{\dagger} +\frac{\hbar^2}{2m} \partial^2_x \WF^{\dagger} - V \WF^{\dagger} =  (\d  - \WF) ,
	\label{eq.1D_Schroedinger_adjoint_equation}
\end{flalign}
where we have called the Lagrange multiplier $\Lm$, as the time-reversed adjoint wavefunction $\Lm=\WF^{\dagger}$ in order to satisfy boundary conditions.

\subsubsection{1D Fr\'echet derivatives}

If the dependence of $\WF^{\dagger}$ on the model is ignored, i.e., we assume that the wavefunction $\WF^{\dagger}$ perfectly satisfies the adjoint equation of motion eq. \eqref{eq.1D_Schroedinger_adjoint_equation} and perturbations in the model parameters exist $(\delta\hbar\neq0,\ \delta m\neq0,\ \delta V\neq0)$, we obtain 
\begin{align}
	\begin{aligned}
		\delta_\m \chi &= \left \langle  \WF^{\dagger} , \left[\imag \, \delta \hbar \, \partial_t \WF + \frac{\hbar^2}{2m^2} \delta m \, \partial^2_x \WF -
		\delta V\,\WF \right] \right \rangle^c_{x,t} , \\
		& = \left \langle \delta \ln \hbar , K_{\hbar} \right \rangle_{x} + \left \langle \delta \ln m , K_{m} \right \rangle_{x} +
		\left\langle \delta\ln V,K_V\right\rangle_x = 0, 
	\end{aligned}
	\label{eq.1D_Schroedinger_variation_action_kernels}
\end{align}
where we have defined the following Fr\'echet sensitivity kernels
\begin{align}
	K_{\hbar}  = \imag \, \hbar \left \langle  \WF^{\dagger} , \partial_t\WF \right \rangle^c_{t} , \qquad 	K_{m}  = \frac{\hbar^2}{2m}\left \langle  \WF^{\dagger} , \partial_x^2\WF \right \rangle^c_{t} \qquad 
	K_V
	&=
	-V
	\left\langle
	\WF^{\dagger},\WF
	\right\rangle_t^c .
	\label{eq.1D_Schroedinger_Adjoint_Kernels}
\end{align}

We can observe that the Fr\'echet interaction density $K_R$ (eq. \eqref{eq.Frechet_interaction}) works as a fundamental bilinear quantity, the instantaneous overlap between forward and adjoint wavefields, from which the Fr\'echet kernels defined in eq. \eqref{eq.1D_Schroedinger_Adjoint_Kernels} are constructed. In particular, the sensitivity kernel with respect to the potential $V(x)$ is directly proportional to the time-integrated Fr\'echet interaction density,
\begin{align}
	K_V
	=
	-V\int_T K_R\,dt
	=
	-V K_R .
\end{align}
Thus, perturbations of the potential provide the most direct connection between the Fr\'echet interaction density and the sensitivity of the Schr\"odinger equation to its model parameters.

A direct probabilistic interpretation arises only for a particular choice of the adjoint field. When $\WF^{\dagger}=\overline{\WF}$, the Fr\'echet interaction density reduces to $|\WF|^2$, which is real and non-negative, and can be interpreted, after normalization, as a probability density. This illustrates how the probability structure emerges as a specific realization of the more general sensitivity kernel defined by the variational formulation.

\section{Discussion}

\subsection{The Fr\'echet interaction density and the emergence of probability}

The forward wavefunction $\WF$ describes the evolution of the quantum system from its initial conditions according to the Schr\"odinger equation. On its own, however, it does not specify a definite measurement outcome. The adjoint wavefunction $\WF^{\dagger}$ has a different role. It propagates sensitivity backward in time from a chosen measurement functional or observable and provides the complementary field required by the variational formulation.

Mathematically, $\WF^{\dagger}$ arises from the constraint of the variational problem. In the interpretation proposed here, it acts as a \emph{sensitivity probe}, describing how different regions of space--time contribute to a chosen observable. The forward and adjoint wavefunctions thus describe two different aspects of the problem: the evolution of the quantum state and its sensitivity to the chosen measurement functional.

The central object connecting these two quantities is the Fr\'echet interaction density (eq.~\eqref{eq.Frechet_interaction}), defined by the local interaction between the forward and adjoint wavefunctions. In general, this interaction defines a sensitivity density and may be complex-valued. For the particular choice $\WF^{\dagger}=\overline{\WF}$, it becomes
\begin{align}
	K_R(\x,t)
	=
	\overline{\WF}(\x,t)\WF(\x,t)
	=
	|\WF(\x,t)|^2,
\end{align}
which is real and non-negative and coincides, after normalization, with the Born probability density. The Born probability density can thus be understood as a particular realization of the more general Fr\'echet interaction density.

This provides a direct connection between the variational sensitivity structure and the probabilistic structure of quantum mechanics. The general interaction between forward and adjoint wavefunctions describes sensitivity to a chosen observable, while the particular choice $\WF^{\dagger}=\overline{\WF}$ gives the positive density associated with probability.

\subsection{Interpretation of the residual functional $\error_op$}

The general misfit/residual functional $\error_op$ defined in eq.~\eqref{eq.quantum_chi_squared} measures a discrepancy between data $(\d)$ and a modeled wavefunction $(\WF)$. In classical adjoint inverse problems, this difference is usually defined with respect to external observations. In quantum mechanics, however, the wavefunction is not directly observed and measurement outcomes are probabilistic. As a result, the notion of a residual functional must be interpreted in a more general way.

In the presented variational formulation, $\error_op$ may represent a \emph{measurement functional} or an \emph{observable}, rather than only a direct comparison with external data. The adjoint wavefunction is associated with this functional and describes how the chosen quantity responds to perturbations in the wavefunction. Different choices of $\error_op$ in eq.~\eqref{eq.quantum_chi_squared} lead to different adjoint sources in eq. \eqref{eq.General_Adjoint_Equation}, and may therefore produce different Fr\'echet sensitivity kernels. The adjoint wavefunction is not unique, but depends on the functional considered in the variational problem.

The Born probability density requires the particular choice $\WF^\dagger=\overline{\WF}$. In terms of the adjoint equation (eq.~\eqref{eq.General_Adjoint_Equation}), this corresponds to choosing an adjoint source consistent with this solution. The Born rule is thus not uniquely determined by the variational formulation, but appears as a particular case of the more general Fr\'echet interaction density.

Within this interpretation, different choices of the functional may lead to different sensitivity kernels. The standard probabilistic case is obtained when the adjoint choice gives a real, non-negative, and normalizable interaction density.

\subsection{The Fr\'echet interaction density and quantum observables} 

Since the Fr\'echet interaction density (eq.~\eqref{eq.Frechet_interaction}) corresponds to the quadratic overlap between forward and adjoint wavefunctions and when the adjoint is chosen as the complex conjugate of the forward field, $\WF^{\dagger} = \overline{\WF}$, the integrand reduces to $|\WF|^2$, which coincides with the standard probability density in quantum mechanics.
	
More generally, observable quantities can be expressed as bilinear forms involving the forward wavefunction and an operator acting on it (eq. \eqref{eq.General_Frechet_Kernel}). Within the present framework, this corresponds to applying linear operators to the forward wavefield within the Fr\'echet sensitivity structure. This yields a family of observable kernels of the form
\begin{itemize}
	\item[\ding{226}] Probability density 
	\begin{equation}
		K_{\rho}(\x,t) = |\WF(\x,t)|^2
	\end{equation}
	
	\item[\ding{226}] Momentum density $(\widehat{p}=-\imag \, \hbar \, \partial_\x)$
	\begin{equation}
		K_{\widehat{p}}(\x,t) = \overline{\WF}(\x,t)\,\widehat{p}\,\WF(\x,t)
		= -\imag \, \hbar \, \overline{\WF}(\x,t)\,\partial_\x \WF(\x,t)
	\end{equation}
	
	\item[\ding{226}] Energy density
	\begin{equation}
		K_E(\x,t) = \overline{\WF}(\x,t)\,\widehat{H}\,\WF(\x,t)
	\end{equation}
	
	\item[\ding{226}] General observable $\widehat{A}$
	\begin{equation}
		K_{\widehat{A}}(\x,t) = \overline{\WF}(\x,t)\,\widehat{A}\,\WF(\x,t)
	\end{equation}
\end{itemize}

These expressions show that quantum observables naturally arise as bilinear forms within the same forward--adjoint structure that defines the Fr\'echet kernel. The fr\'echet interaction density corresponds to the special case in which no operator is applied, yielding a positive-definite quantity that can be normalized and interpreted probabilistic.
	
\subsection{Variational adjoints and the two-state vector formalism}

Yakir Aharonov and colleagues proposed the Two-State Vector Formalism (TSVF) \citep{aharonov1964time}, where a quantum system is described not only by a forward-evolving state from an initial condition, but also by a backward-evolving state from a final post-selection. In TSVF, both past and future boundary conditions are treated as physically real, jointly determining the system's evolution in between. Therefore, the reality of the system is entangled with its future outcomes. This means that the future measurements influence the conditional probabilities of past events, but this does not imply signaling backward in time. The existence of a particle at intermediate times is contingent upon both its initial preparation and the post-selection outcome. This conditional structure is a hallmark of the TSVF perspective since it explicitly allows arbitrary backward states and interprets them as a physically post-selected state.

The variational--adjoint interpretation shares a conceptual similarity with the TSVF, since both involve forward- and backward-evolving wavefunctions. In the present interpretation, however, the backward-propagating field arises from the variational structure of the Schr\"odinger equation and describes the sensitivity of a chosen measurement functional or observable to the forward wavefunction. Unlike the backward state in the TSVF, the adjoint wavefunction is not introduced as a physically post-selected quantum state. It first appears as the dual variable of the variational problem, to which we associate a physical interpretation in terms of measurement sensitivity.

The interaction between the forward and adjoint wavefunctions then describes how different regions of space--time contribute to the chosen observable. In this sense, the variational--adjoint interpretation can be summarized as
\[
\text{forward evolution + backward sensitivity = measurement information.}
\]
The similarity with the TSVF lies in the use of forward and backward quantities, while their origin and interpretation are different. In the TSVF, the backward-evolving state is determined by a physical post-selection; in the present formulation, the backward-evolving adjoint is determined by the variational problem and the chosen functional. The adjoint formulation then provides a systematic way of constructing Fr\'echet sensitivity kernels, with the Born probability density obtained for the particular choice $\WF^\dagger=\overline{\WF}$.

\subsection{Implications for quantum control}

The variational formulation developed here has a natural connection with quantum optimal control. In quantum control, the main goal is to modify the evolution of a quantum system in order to reach a desired dynamical or final state. Optimal control methods \citep{peirce1988optimal,khaneja2005optimal,brif2010control} commonly use adjoint equations to calculate how variations in control fields or Hamiltonian parameters affect a chosen objective functional.

The Fr\'echet sensitivity kernels introduced here have the same mathematical role. They quantify how perturbations in the parameters of the Schr\"odinger equation affect a chosen functional or observable. The adjoint wavefunction provides the information required to compute these sensitivities efficiently and can therefore be used to construct gradients with respect to control parameters. This connection places the present formulation close to the mathematical structure used in optimal control and Pontryagin's maximum principle
\citep{mangasarian1966sufficient,pontryagin1962mathematical}.

There is, however, an important difference in interpretation. In quantum optimal control, the adjoint wavefunction is mainly introduced as a mathematical tool for calculating gradients and optimizing control fields. In the interpretation proposed here, we also associate the adjoint wavefunction with measurement sensitivity. Its interaction with the forward wavefunction defines the Fr\'echet interaction density and, for the particular choice $\WF^\dagger=\overline{\WF}$, gives the Born probability density.

The same mathematical structure is also relevant to quantum measurement and metrology. Different measurement functionals produce different adjoint sources and sensitivity kernels, allowing one to study how a chosen observable responds to variations in the wavefunction, Hamiltonian parameters, or control fields. Such sensitivity information is relevant to problems involving parameter estimation, measurement design, and quantum metrology \citep{wiseman2009quantum,jacobs2014quantum,braunstein1994statistical,paris2009quantum,degen2017quantum}.

The connection with quantum control is both mathematical and conceptual. The same forward--adjoint structure used to calculate gradients in optimal control provides, in the present interpretation, a general description of sensitivity in quantum systems. This suggests possible applications of the Fr\'echet sensitivity kernels to quantum control, measurement, and metrology.

\subsection{On the origin of time-reversed dynamics and the Born rule}

It is important to emphasize that no explicit assumption of backward-in-time physical propagation is introduced in the present framework. The adjoint wavefield arises naturally from the variational formulation as the dual variable enforcing the Schr\"odinger constraint. Its apparent backward-in-time evolution is a direct consequence of the adjoint operator structure, which propagates sensitivity from the measurement functional to earlier times. 

In contrast to stochastic approaches \citep[e.g.][]{nelson1966derivation}, where backward dynamics must be introduced to restore time symmetry, the adjoint formulation produces this dual evolution intrinsically. The forward wavefield describes the propagation of the quantum state, while the adjoint wavefield encodes how a given observable depends on that evolution. The distinction is therefore not one of physical asymmetry, but of mathematical duality between state propagation and sensitivity.

It is also important to clarify that the Born rule is not uniquely derived from the present formulation. The variational formulation defines a general Fr\'echet sensitivity kernel
$K=\WF^\dagger\WF$, which can in general be complex-valued and cannot always be interpreted as a probability density. For such an interpretation, the kernel must be real, non-negative, and normalizable. These properties are obtained for the particular choice $\WF^\dagger=\overline{\WF}$, for which the kernel becomes $K=|\WF|^2$.

In this sense, the Born rule is not introduced as an external postulate, \textit{but neither} is it uniquely determined by the variational formulation. It appears as a particular choice within the more general class of sensitivity kernels, for which the kernel becomes real, non-negative, and normalizable as required for a probability density.

\subsection{Apparent asymmetry between forward and adjoint wavefunctions}

In the presented variational--adjoint interpretation, there is an apparent time asymmetry between the forward and adjoint wavefunctions. This asymmetry naturally arises from their different roles in the variational formulation of the Schr\"odinger equation.

The forward wavefunction $\WF$ evolves according to the Schr\"odinger equation and describes the quantum state of the system. Its evolution is determined by the Hamiltonian and does not, by itself, select a particular measurement outcome.

The adjoint wavefunction $\WF^\dagger$, on the other hand, is associated with a measurement functional or observable of interest and evolves according to the adjoint Schr\"odinger equation. Its role is to quantify how perturbations in $\WF$ affect the chosen functional. Mathematically, $\WF^\dagger$ arises from the constraint imposed in the variational problem; in the interpretation proposed here, it identifies the regions of space--time that are relevant to a particular observable.

This construction is analogous to classical quantum optimal control \citep[e.g.][]{peirce1988optimal,khaneja2005optimal,brif2010control}, where the forward state evolves according to the governing equation, while the adjoint describes the sensitivity of a cost functional to variations in the state. In the present interpretation, the measurement functional plays a similar role, and $\WF^\dagger$ acts as a sensitivity probe that describes how changes in $\WF$ affect the observable.

The Fr\'echet interaction density combines these two wavefunctions and quantifies their local interaction. For the particular choice $\WF^\dagger=\overline{\WF}$, the interaction density reduces to $|\WF|^2$, giving the Born probability density. The apparent asymmetry between $\WF$ and $\WF^\dagger$ comes from their different roles: $\WF$ describes the evolution of the quantum system, while $\WF^\dagger$ describes its sensitivity with respect to the chosen measurement functional. Probabilities and other observable kernels can
then be associated with their interaction through the Fr\'echet interaction density.

\section{Conclusions}

We have presented a variational interpretation of quantum mechanics in which the interaction between forward and adjoint wavefunctions defines a general Fr\'echet sensitivity kernel which encodes how different regions of space--time contribute to a chosen observable and/or functional. In general, this kernel is complex-valued and may take negative or
imaginary values, and cannot be directly interpreted as a probability density. It should instead be understood as a sensitivity or influence field, analogous to adjoint kernels in classical wave physics, highlighting where the system is most responsive to perturbations.

Within this interpretation, the Born rule appears as a particular case. When the adjoint wavefunction is identified with the complex conjugate of the forward wavefunction,
$\WF^{\dagger}=\overline{\WF}$, consistently satisfying the adjoint Schr\"odinger equation, the Fr\'echet interaction density reduces to
\begin{align}
	K_R = \overline{\WF}\WF = |\WF|^2 ,
\end{align}
which is real and non-negative. This particular identification provides a direct connection between the variational sensitivity structure and the Born probability density. Other choices of $\WF^\dagger$ do not, in general, give a physically interpretable probability density.

It is important to emphasize the conceptual difference between the use of adjoint fields in quantum control and in the adjoint interpretation proposed here. In quantum control, adjoint wavefunctions are used to calculate gradients of a cost functional efficiently, determining how variations in control fields or Hamiltonian parameters affect a chosen target. The physical interpretation of the adjoint wavefunction is typically not the main interest, since it serves primarily as an optimization tool.

In the interpretation proposed here, we give the adjoint wavefunction $\WF^\dagger$ an additional physical meaning associated with the sensitivity of the system to measurement. Its interaction with the forward wavefunction $\WF$ defines observable kernels, with the Born probability density obtained for the particular choice $\WF^\dagger=\overline{\WF}$. In this sense, the forward wavefunction describes the evolution of the quantum system, while the adjoint wavefunction provides a variational representation of measurement sensitivity.

\section{Acknowledgments}

The author gratefully acknowledge an anonymous reviewer for very constructive comments that helped to improve the manuscript.

\section{Conflict of interest}

The authors declare no conflict of interest.

\footnotesize
\bibliographystyle{apalike}
\bibliography{Biblio}

@article{tonti1973variational,
  title={On the variational formulation for linear initial value problems},
  author={Tonti, Enzo},
  journal={Annali di Matematica Pura ed Applicata},
  volume={95},
  number={1},
  pages={331--359},
  year={1973},
  publisher={Springer}
}

@book{griffel2002applied,
	title={Applied Functional Analysis},
	author={Griffel, David},
	year={1981},
	publisher={Wiley}
}

@book{boyd2004convex,
  title={Convex Optimization},
  author={Boyd, Stephen and Boyd, Stephen P and Vandenberghe, Lieven},
  year={2004},
  publisher={Cambridge University Press}
}

@book{zeidler1984nonlinear,
	title={{Nonlinear functional analysis and its applications III: variational methods and optimization}},
	author={Zeidler, Eberhard},
	year={1984},
	publisher={Springer}
}

@book{zeidler1990nonlinear,
	title={{Nonlinear functional analysis and its applications II: linear monotone operators}},
	author={Zeidler, Eberhard},
	year={1990},
	publisher={Springer}
}

@article{abreu2024understanding,
	title={Understanding the Adjoint Method in Seismology: Theory and Implementation in the Time Domain},
	author={Abreu, Rafael},
	journal={Surveys in Geophysics},
	volume={45},
	number={5},
	pages={1363--1434},
	year={2024},
	publisher={Springer}
}

@article{hanson1981sufficiency,
  title={{On sufficiency of the Kuhn-Tucker conditions}},
  author={Hanson, Morgan A},
  journal={Journal of Mathematical Analysis and Applications},
  volume={80},
  number={2},
  pages={545--550},
  year={1981}
}

@article{hanson1999invexity,
	title={{Invexity and the Kuhn--Tucker theorem}},
	author={Hanson, Morgan A},
	journal={Journal of Mathematical Analysis and Applications},
	volume={236},
	number={2},
	pages={594--604},
	year={1999},
	publisher={Elsevier}
}

@book{griffiths2018introduction,
	title={Introduction to quantum mechanics},
	author={Griffiths, David J and Schroeter, Darrell F},
	year={2018},
	publisher={Cambridge University Press}
}

@book{shankar2012principles,
	title={Principles of quantum mechanics},
	author={Shankar, Ramamurti},
	year={2012},
	publisher={Springer}
}

@book{zee2010quantum,
	title={Quantum field theory in a nutshell},
	author={Zee, Anthony},
	volume={7},
	year={2010},
	publisher={Princeton University Press}
}

@book{sakurai2020modern,
	title={Modern quantum mechanics},
	author={Sakurai, Jun John and Napolitano, Jim},
	year={2020},
	publisher={Cambridge University Press}
}

@book{Feynman1965,
	author    = {Richard P. Feynman and Robert B. Leighton and Matthew Sands},
	title     = {The Feynman Lectures on Physics, Volume III: Quantum Mechanics},
	publisher = {Addison-Wesley},
	address   = {Reading, MA},
	year      = {1965}
}

@article{cramer1986transactional,
	author    = {John G. Cramer},
	title     = {The Transactional Interpretation of Quantum Mechanics},
	journal   = {Reviews of Modern Physics},
	volume    = {58},
	number    = {3},
	pages     = {647--688},
	year      = {1986},
	doi       = {10.1103/RevModPhys.58.647}
}

@book{cramer2016quantum,
	title={The quantum handshake},
	author={Cramer, John G},
	year={2016},
	publisher={Springer}
}

@book{bohr1935copenhagen,
	author    = {Niels Bohr},
	title     = {Atomic Physics and Human Knowledge},
	publisher = {Wiley},
	year      = {1958},
	note      = {Contains Bohr's essays outlining the Copenhagen interpretation}
}

@article{Everett1957,
	title = {"Relative State" Formulation of Quantum Mechanics},
	author = {Everett, Hugh},
	journal = {Reviews of Modern Physics},
	volume = {29},
	issue = {3},
	pages = {454--462},
	year = {1957}
}

@article{DeWitt1970,
	author = {DeWitt, Bryce S.},
	title = {Quantum mechanics and reality},
	journal = {Physics Today},
	volume = {23},
	number = {9},
	pages = {30-35},
	year = {1970},
	month = {09},
	doi = {10.1063/1.3022331}
}

@article{Bohm1,
	title={{A suggested interpretation of the quantum theory in terms of "hidden" variables. I}},
	author={Bohm, David},
	journal={Physical Review},
	volume={85},
	number={2},
	pages={166},
	year={1952},
	publisher={APS}
}

@article{Bohm2,
	title = {{A suggested interpretation of the quantum theory in terms of "hidden" variables. II}},
	author = {Bohm, David},
	journal = {Physical Review},
	volume = {85},
	issue = {2},
	pages = {180--193},
	numpages = {0},
	year = {1952},
	month = {Jan},
	publisher = {American Physical Society},
	doi = {10.1103/PhysRev.85.180}
}

@book{Holland_1993, 
	title={{The Quantum Theory of Motion: An Account of the de Broglie-Bohm Causal Interpretation of Quantum Mechanics}}, 
	publisher={Cambridge University Press}, 
	author={Holland, Peter R.}, 
	year={1993}}

@article{Ghirardi1986,
	title = {Unified dynamics for microscopic and macroscopic systems},
	author = {Ghirardi, G. C. and Rimini, A. and Weber, T.},
	journal = {Physcal Review D},
	volume = {34},
	issue = {2},
	pages = {470--491},
	numpages = {0},
	year = {1986},
	month = {Jul},
	publisher = {American Physical Society},
	doi = {10.1103/PhysRevD.34.470}
}

@article{griffiths1984consistent,
	title={Consistent histories and the interpretation of quantum mechanics},
	author={Griffiths, Robert B},
	journal={Journal of Statistical Physics},
	volume={36},
	number={1},
	pages={219--272},
	year={1984},
	publisher={Springer}
}

@book{griffiths2003consistent,
	title={Consistent quantum theory},
	author={Griffiths, Robert B},
	year={2003},
	publisher={Cambridge University Press}
}

@article{fuchs2014introduction,
	title={{An introduction to QBism with an application to the locality of quantum mechanics}},
	author={Fuchs, Christopher A and Mermin, N David and Schack, R{\"u}diger},
	journal={American Journal of Physics},
	volume={82},
	number={8},
	pages={749--754},
	year={2014},
	publisher={AIP Publishing}
}

@article{rovelli1996relational,
	title={Relational quantum mechanics},
	author={Rovelli, Carlo},
	journal={International Journal of Theoretical Physics},
	volume={35},
	number={8},
	pages={1637--1678},
	year={1996},
	publisher={Springer}
}

@article{aharonov1964time,
	title={Time symmetry in the quantum process of measurement},
	author={Aharonov, Yakir and Bergmann, Peter G and Lebowitz, Joel L},
	journal={Physical Review},
	volume={134},
	number={6B},
	pages={B1410},
	year={1964},
	publisher={APS}
}

@book{nielsen2010quantum,
	title={Quantum computation and quantum information},
	author={Nielsen, Michael A and Chuang, Isaac L},
	year={2010},
	publisher={Cambridge University Press}
}

@article{dowling2003quantum,
	title={Quantum technology: the second quantum revolution},
	author={Dowling, Jonathan P and Milburn, Gerard J},
	journal={Philosophical Transactions of the Royal Society of London. Series A: Mathematical, Physical and Engineering Sciences},
	volume={361},
	number={1809},
	pages={1655--1674},
	year={2003},
	publisher={The Royal Society}
}

@book{ballentine2014quantum,
	title={Quantum mechanics: a modern development},
	author={Ballentine, Leslie E},
	year={2014},
	publisher={World Scientific Publishing Company}
}

@article{abreu2025frechetrootkernelcertain,
	title={{On the Fr\'echet interaction density of certain wave equations}},
	author={Rafael Abreu and Chahana Nagesh},
	journal={\href{https://arxiv.org/abs/2512.09609}{\color{black} Preprint arXiv:2512.09609}},
	year={2025}
}

@article{peirce1988optimal,
	title={Optimal control of quantum-mechanical systems: Existence, numerical approximation, and applications},
	author={Peirce, Anthony P. and Dahleh, Munther A. and Rabitz, Herschel},
	journal={Physical Review A},
	volume={37},
	number={12},
	pages={4950--4964},
	year={1988},
	publisher={APS}
}

@article{khaneja2005optimal,
	title={{Optimal control of coupled spin dynamics: design of NMR pulse sequences by gradient ascent algorithms}},
	author={Khaneja, Navin and Reiss, Tobias and Kehlet, Cindie and Schulte-Herbr{\"u}ggen, Thomas and Glaser, Steffen J.},
	journal={Journal of Magnetic Resonance},
	volume={172},
	number={2},
	pages={296--305},
	year={2005},
	publisher={Elsevier}
}

@article{brif2010control,
	title={Control of quantum phenomena: past, present and future},
	author={Brif, Constantin and Chakrabarti, Raj and Rabitz, Herschel},
	journal={New Journal of Physics},
	volume={12},
	number={7},
	pages={075008},
	year={2010},
	publisher={IOP Publishing}
}

@book{wiseman2009quantum,
	title={Quantum Measurement and Control},
	author={Wiseman, Howard M. and Milburn, Gerard J.},
	year={2009},
	publisher={Cambridge University Press}
}

@book{jacobs2014quantum,
	title={Quantum Measurement Theory and its Applications},
	author={Jacobs, Kurt},
	year={2014},
	publisher={Cambridge University Press}
}

@article{braunstein1994statistical,
	title={Statistical distance and the geometry of quantum states},
	author={Braunstein, Samuel L. and Caves, Carlton M.},
	journal={Physical Review Letters},
	volume={72},
	number={22},
	pages={3439--3443},
	year={1994},
	publisher={APS}
}

@article{paris2009quantum,
	title={Quantum estimation for quantum technology},
	author={Paris, Matteo G. A.},
	journal={International Journal of Quantum Information},
	volume={7},
	number={supp01},
	pages={125--137},
	year={2009},
	publisher={World Scientific}
}

@article{degen2017quantum,
	title={Quantum sensing},
	author={Degen, Christian L. and Reinhard, Friedemann and Cappellaro, Paola},
	journal={Reviews of Modern Physics},
	volume={89},
	number={3},
	pages={035002},
	year={2017},
	publisher={APS}
}

@article{mangasarian1966sufficient,
	title={Sufficient conditions for the optimal control of nonlinear systems},
	author={Mangasarian, Olvi L},
	journal={SIAM Journal on control},
	volume={4},
	number={1},
	pages={139--152},
	year={1966},
	publisher={SIAM}
}

@article{nelson1966derivation,
	title={{Derivation of the Schr{\"o}dinger equation from Newtonian mechanics}},
	author={Nelson, Edward},
	journal={Physical Review},
	volume={150},
	number={4},
	pages={1079},
	year={1966},
	publisher={APS}
}

@article{caticha2011entropic,
	title={Entropic dynamics, time and quantum theory},
	author={Caticha, Ariel},
	journal={Journal of Physics A: Mathematical and Theoretical},
	volume={44},
	number={22},
	pages={225303},
	year={2011}
}

@article{hall2002schrodinger,
	title={{Schr{\"o}dinger equation from an exact uncertainty principle}},
	author={Hall, Michael JW and Reginatto, Marcel},
	journal={Journal of Physics A: Mathematical and General},
	volume={35},
	number={14},
	pages={3289--3303},
	year={2002}
}

@article{yang2024quantum,
	title={Quantum mechanics based on an extended least action principle and information metrics of vacuum fluctuations},
	author={Yang, Jianhao M},
	journal={Foundations of Physics},
	volume={54},
	number={3},
	pages={32},
	year={2024},
	publisher={Springer}
}

@article{stoica2025born,
	title={Born rule: quantum probability as classical probability},
	author={Stoica, Ovidiu Cristinel},
	journal={International Journal of Theoretical Physics},
	volume={64},
	number={5},
	pages={117},
	year={2025},
	publisher={Springer}
}

@article{zhang2026schrodinger,
	title={{Derivation of the Schr\"odinger equation from fundamental principles}},
	author={Zhang, Wenzhuo and Svidzinsky, Anatoly A.},
	journal={Encyclopedia},
	volume={6},
	number={2},
	pages={41},
	year={2026},
	publisher={MDPI}
}

@article{neumaier2025born,
	title={{The Born rule—100 years ago and today}},
	author={Neumaier, Arnold},
	journal={Entropy},
	volume={27},
	number={4},
	pages={415},
	year={2025},
	publisher={MDPI}
}

@book{saunders2010many,
	title={{Many worlds?: Everett, quantum theory, \& reality}},
	author={Saunders, Simon},
	year={2010},
	publisher={Oxford University Press}
}

@book{neumaier2019coherent,
	title={Coherent quantum physics: a reinterpretation of the tradition},
	author={Neumaier, Arnold},
	year={2019},
	publisher={Walter de Gruyter GmbH \& Co KG}
}

@book{pontryagin1962mathematical,
  title     = {The Mathematical Theory of Optimal Processes},
  author    = {Pontryagin, L. S. and Boltyanskii, V. G. and Gamkrelidze, R. V. and Mishchenko, E. F.},
  year      = {1986},
  publisher = {Gordon and Breach Science Publishers}
}

\end{document}